\documentclass[pra,reprint,flushbottom,showpacs]{revtex4-1}
\usepackage{epsfig}
\usepackage{amsmath}
\usepackage{amssymb}
\usepackage{amsthm} 
\usepackage{bm}
\usepackage{color}

\newcommand{\ket}[1]{\left|#1\right>}
\newcommand{\bra}[1]{\left<#1\right|}
\newcommand{\nn}{\nonumber\\}

\newcommand{\bea}{\begin{eqnarray}}
\newcommand{\ea}{\end{eqnarray}}
\newcommand{\eea}{\end{eqnarray}}
\newcommand{\ord}{{\cal O}}
\newcommand{\sumint}[1]

\begin{document}

\title{Many-site coherence revivals in the extended Bose-Hubbard model \\
and the Gutzwiller approximation} 
\author{Uwe R. Fischer and Bo Xiong}

\affiliation{Seoul National University,   Department of Physics and Astronomy \\  Center for Theoretical Physics, 
151-747 Seoul, Korea}

\begin{abstract}
We investigate the collapse and revival of first-order coherence in deep optical lattices when long-range interactions are turned on, 
and find that the first few revival peaks are strongly attenuated already for moderate values
of the nearest-neighbor interaction coupling. It is shown that the conventionally employed 
Gutzwiller wavefunction, with only onsite-number dependence of the variational amplitudes, leads to incorrect predictions for the collapse and revival oscillations within the extended 
Bose-Hubbard model. We provide a modified variant of the Gutzwiller ansatz, reproducing the analytically calculated time dependence of first-order coherence in the limit of zero tunneling. 
\end{abstract}


\maketitle

\section{Introduction} 
The Hubbard model describes interacting particles 
in sufficiently deep periodic lattice potentials, so that the tight-binding approach applies, and the Wannier functions form the basis of the bosonic or fermionic 
field operator expansion \cite{Gutzwiller,Hubbard,Kanamori}. 
It was realized about a decade ago that in optical lattices, generated by
counterpropagating laser beams, a clean and highly controllable implementation
of the Hubbard model can be obtained 
\cite{Jaksch}.  
Initially, attempts to achieve a precise 
understanding of the collective behavior of atoms or molecules trapped in the lattice
was concentrated on the classic version of the model with onsite interactions only. 
However, a zoo of quantum phases beyond the superfluid and Mott phases, like various 
supersolid and charge-density wave phases, emerges when long-range interactions are 
turned on \cite{Trefzger}. The most prominent and (for electrically neutral atoms/molecules) physically realizable examples of the latter 
are magnetic and electric dipole-dipole interactions \cite{Baranov}. The 
so-called extended Hubbard model which incorporates the long-range character 
of the interactions \cite{history}.
It is crucial for the determination of possible phases to obtain information on the interaction parameter values in the extended Bose-Hubbard model, and we suggest, in the following, experimental means to measure them by using the temporal variations of first-order coherence. 


Collapse and revival oscillations are familiar from quantum optics \cite{Robinett} 
as a sensitive measure for the spectrum of the Hamiltonian 
 and the many-body dynamics of the system. 
In ultracold bosonic quantum gases, collapse and revival oscillations were experimentally observed
both on the two- and multi-body interaction level \cite{Greiner,Will}, when switching quasi-instantaneously from the superfluid to the Mott side in a dynamical quantum phase transition, a quench \cite{Dziarmaga}. 
Rapid experimental progress in the meantime even allows for the single-site and single-atom resolution of the time evolution of the site occupation number \cite{Bakr,Weitenberg}.
  
In the optical lattice realization of collapse and revival of coherence,  
effectively many small samples on the various sites are prepared, which are 
{\em independent} of each other when only contact interactions are present. When the
particle numbers in individual sites are small, i.e. at small filling, 
the revival times are short enough for observing the collapse and revival  phenomenon \cite{Wright}. The collapse and revival times have in addition been demonstrated to be sensitive to the effective dimension of a single well and its trap power law \cite{You}.  
  
It has been previously shown that a sensitive measure of the many-body dynamics
due to collapse and revival dynamics in optical lattices may be experimentally accessible when tunneling across different sites is included \cite{Fischer,Wolf,Schachenmayer}. Furthermore, significant effects on collapse and revival can be expected 
when effective three-body interactions \cite{Johnson} as well as the number squeezing 
of a initial state with finite interactions on the superfluid side are taken into account \cite{Tiesinga}. In addition, the  number squeezing of an initial state can be caused by 
finite sweep rates, when quenching from the superfluid to the localized Mott phase  \cite{Schuetzhold}.  

We explore in the following effects which are intrinsically beyond the familiar
purely onsite collapse and revival evolution in optical lattices and come from 
long-range interactions (that is 
not those, e.g., coming from an additional external confinement potential \cite{Buchhold}).  
The extended Hubbard model provides nearest-neighbor (NN) coupling of sites
on the two-body interaction level, so that the Hamiltonian 
converts the onsite phenomenon collapse and revival obtained for contact interactions 
into a genuine many-body--many-site effect.  
We demonstrate that an attenuation of the 
first few phase coherence revivals takes place in a range of the  
ratio of onsite and NN interaction couplings 
relevant to the predicted quantum phases of the 
extended Bose-Hubbard model \cite{noteI,Trefzger}. 
It is shown that, as opposed to the reduction of the coherence signal by the 
(phase-sensitive) tunneling term in the Hamiltonian, 
no real damping takes place due to a NN
interaction coupling, instead there are two distinct revival scales related to onsite and NN interaction couplings $U$ and $V$, respectively. The consequent 
effective attenuation  of the 
first few revival peaks due to $V$ is much stronger than those to tunneling damping,
increasing with filling rather than decreasing, 
and can easily overcome inhomogenity (external trapping) effects. We therefore propose
a sensitive measure of the value of $V$ realizable in current 
optical lattice experiments.  

The modulation of the first-order coherence signal with frequency $V/\hbar$, 
taking as in Sec. II below a coherent product state as the initial state after a quench into 
a deep lattice, does not occur in the conventional time-dependent Gutzwiller approach.
This approach employs a product state of purely onsite-oocupation dependent variational wavefunctions, and is indeed widely used for ground-state calculations in the extended Bose-Hubbard model, cf. e.g. \cite{Trefzger,Goral,Scarola,Iskin}. As we will show below, it however leads at least in dynamical situations (i.e. with explicit time dependence of correlation functions) 
to in general grossly incorrect predictions and hence needs modification. 
We will devise a variant of the Gutzwiller product state with the variational 
amplitudes depending on the NN particle-numbers. The form of this wavefunction is based 
on the analytical evolution of first-order phase coherence in deep optical lattices, 
i.e., when tunneling is exponentially small and thus negligible. 
We then apply this variant of the Gutzwiller wavefunction, appropriate to the extended
Bose-Hubbard model, to an initial state which is number-squeezed. We again compare 
the result with that obtained by the conventional Gutzwiller approach, showing that
in most experimentally relevant cases the NN-coupling-induced attenuation of the 
coherence signal will completely dominate the damping caused by residual tunneling. 

\section{coherent initial  state} 
The Hamiltonian of the single-band extended Bose-Hubbard model 
is assumed to be of the form  
\bea
\hat H &=& - J  \sum_{<ij>}^M \hat a_i^\dagger \hat a_j  
+ \frac U2 \sum_i^M (\hat n_i-1) \hat n_i 
+ \frac V2 \sum_{<ij>}^M \hat n_i \hat n_j \nn
\label{H}
\ea 
where $<\!ij\!>$ indicates summation over NN, and $M$ is the number of lattice sites. 
We consider here the most general situation that the parameters $J,U,V$
can be controlled  independently of each other, e.g. 
by optical lattice depth and the ratio of contact and long-range interactions. 
Their values are determined by overlap integrals of Wannier functions incorporating   
single-particle energy operators and two-body interaction power laws and effective ranges,
respectively. 
The two-body interaction terms correspond to density-density
type interactions with onsite ($U$) and NN ($V$) parts.

We take in this section the initial state to be the product of onsite coherent (superfluid) states with Poissonian probability distribution of site occupation  
\bea
\ket{\rm SF}
= 
\prod_{i=1}^M e^{-|\alpha_i^2|/2} 
\sum_{n_i=0}^\infty \frac{\alpha_i^{n_i}}{\sqrt{n_i !}}\ket{n_i} \label{coh} 
\ea
where $\ket{n_i}$ are local number eigenstates,
$\hat n_i \ket{n_i}= n_i \ket{n_i} $
and $\hat a_i \ket{\alpha_i}=\alpha_i \ket{\alpha_i} $
with 
average $ \bar n_i\equiv \langle \hat n_i \rangle=|\alpha_i|^2$.
Such a state is for example physically realizable with a quasi-instantaneous sweep across the superfluid--Mott transition starting from (deep inside) 
the superfluid phase, in which $J\gg U, V$ \cite{Greiner,Fischer}. 
We note that finite $M$ corrections due to a 
particle-number-conserving initial state and the resulting time evolution
were investigated for the Bose-Hubbard model in [18]. However, they became negligible for a number of sites $M\gtrsim 5$, so that we can expect to be able to safely neglect them for sufficiently large $M$ also with nonvanishing off-site interaction coupling.


The limits of either $U$ and $J$ being dominant over the other couplings are
well known, and lead to the (deep) Mott phase 
(at commensurate filling in the present homogeneous case 
$N/M=\bar n$\,integer, where $N$ is the total number of particles) 
and the superfluid phase of weakly noninteracting bosons, respectively. 
We choose $U\equiv 1$ as the unit 
of energy and $2\pi/U$ as the unit of time ($\hbar \equiv 1$). 

We first derive an analytical result for the exact time evolution 
of first-order coherence when $J=0$ in \eqref{H}. 
Because the energy is expressible in the number (i.e.\,Fock space)
basis when $J=0$, the first order correlation function 
can be analytically determined. 
Calculating with the initial coherent product state \eqref{coh}  
the off-diagonal elements of the single-particle density matrix, one obtains   
\begin{widetext}
\bea
\bra{\rm SF} e^{i\hat H t}  \hat a_i^\dagger 
\hat a_j e^{-i\hat H t} \ket{\rm SF} 
& = &
\prod_{l} 
\exp(-\bar n_l) \sum_{n_l,n_l'}^\infty 
\frac{{\alpha^*_l}^{n_l'}{\alpha_l}^{n_l}}{\sqrt{n_l'!n_l!}}  
\sqrt{n_i' n_j} \delta_{n_i'-1,n_i}\delta_{n_j',n_j-1} \delta_{n_k,n_k'}
\exp\left[it(E_{\{n_l'\}}-E_{\{n_l\}})\right]
\nn
& = &
\alpha^*_i \alpha_j  \prod_{k=i,j,\mbox{NN}\,l} \exp[-\bar n_k] 
\sum_{n_i,n_j=0}^\infty \frac{{\bar n_i}^{n_i}}{n_i!} \frac{{\bar n_j}^{n_j}}{n_j!} 
\sum_{n_l=0}^\infty \frac{{\bar n_l}^{n_l}}{n_l!} 
\exp
\left[
it U(n_i-n_j) + it V (\,\sum_{[il]} n_l - \sum_{[jl]} n_l\, )   
\right]
\nn & = &
\bar n\exp[i\theta_{ij}]
\exp\left[2\bar n (\cos[Ut]-1)\right]
\exp\left[ 2\bar n \nu(\cos[Vt]-1) \right]  
\label{UVrevival}
\ea
\end{widetext} 
where the site-occupation-dependent interaction energy reads 
$E_{\{n_i\}} = \frac U2\sum_i n_i(n_i-1) + \frac V2 \sum_{<ij>} n_i n_j$,
and $[il]$ and $[jl]$ indicate sums over all NNs 
$l$ of $i$ and $j$ at fixed $i$ and $j$, respectively. 
We assume for simplicity in \eqref{UVrevival} that $|i-j|>2$, for which the coherence 
signal becomes independent of distance $|i-j|$ 
(expressions when $|i-j|\le 2$ can easily be derived as well).
The last line is valid for homogeneous filling $\bar n$. 
Finally, $\nu$ is the number of NN, i.e.\,$\sum_{[ik]} =\nu$, 
and $\theta_{ij} = \arg[\alpha_j]-\arg[\alpha_i]$ describes a possible
initial phase offset between sites. 

\begin{center}
\begin{figure}[hbt]
\centering
\includegraphics[width=.36\textwidth]{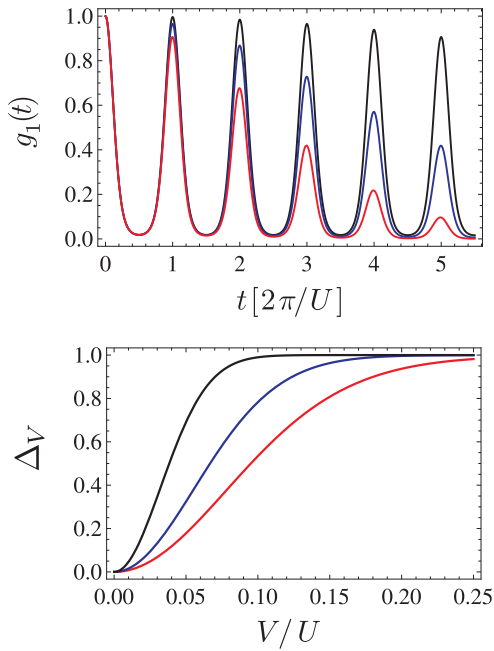}
\vspace*{0.5em}
\caption{\label{Vrole} (Color online) Top: Attenuation of first five revival oscillations due to finite $V/U=0.005,0.015,0.025$ (black, blue, red curves, $V/U$ increasing from top to bottom), according to \eqref{UVrevival},  
$\nu=4$, $\bar n =2$. Bottom: Attenuation factor of first coherence revival $\Delta_V$ in \eqref{DeltaV} as a function of $V/U$, for $\bar n \nu=2,4,12$ (red, blue, black curves, $\bar n \nu$ increasing 
from right to left).}
\end{figure}
\end{center} 
\vspace*{-2em}

The above result contains 
experimentally directly accessible information on the parameters $U$ and $V$   
of the extended Bose-Hubbard model by measuring the quasimomentum distribution function
$n_q (t) = \frac1M \sum_{i,j} \langle \hat a_i^\dagger \hat a_j \rangle e^{iq(x_i-x_j)}$,
whose zero momentum $q=0$ part corresponds to the coherent fraction, where $x_i$ are site centers.
The height of the first few revivals at the instants $2\pi k/U$, where $k$ are positive 
integers, is highly sensitive to the value of $V/U$ (assuming $V<U$),  
cf.\,Fig.\,\ref{Vrole}, where we plot the first order correlation function 
(setting $\theta_i=0$),  
proportional to the zero momentum structure factor $n_0(t)$,
\bea
g_1 (t)  = \frac1{2\bar n} 
\langle \hat a_i^\dagger \hat a_j \rangle + \,{\rm h.c.}
\label{gdef} 
\ea 
A significant attenuation of the revival peaks occurs already for moderate values of $V/U$,
which exceeds the experimentally observed damping 
in the $V\rightarrow 0$ (contact interaction) limit caused, e.g.,
by external trapping and finite temperature \cite{Greiner,Will}. 
 Hence the presently obtained 
 homogeneous system result should readily be observable in 
 experiments like \cite{Will}, which aim at the minimization 
 of external trapping (inhomogenity) effects, and can sensitively detect 
 even small dipole strengths of particles trapped in an optical lattice.

The attenuation of $2\pi/U$ oscillations between revivals at $2\pi/V$ 
increases with the product $\bar n\nu$, cf. Fig.\,\ref{Vrole}.  
We take as a measure of the attenuation the 
value of $\Delta_V=1- \langle \hat a_i^\dagger \hat a_j \rangle /\bar n$ 
at the first onsite revival $t_U=2\pi/U$, 
\bea 
\Delta_V = 1- \exp \left[2\bar n \nu \left(\cos\left[2\pi \frac VU\right]-1\right)\right]\label{DeltaV}
\ea
The overdamped limit corresponds to $\Delta_V\rightarrow 1$.
For small $V/U$, the attenuation is given by $\Delta_V=\bar n \nu (2\pi V/U)^2$.

We now compare our attenuation 
result \eqref{DeltaV} above with the real damping of the revival amplitude due to 
single-particle tunneling coupling between sites. In contrast to the small $J$ result
 $\Delta_J = \left(\frac{2\pi J} U\right)^2 
\left[  \nu(2\nu-1) e^{-4\bar n} I_0^2 (2 \bar n) \right.  
\left. + \nu e^{-3\bar n} F_{\{1,1\}}(\bar n^3) \right]$ (where $I_0,F_{\{1,1\}}$ are modified Bessel and hypergeometric functions, respectively \cite{Fischer}), the 
small V result from \eqref{DeltaV} is linear in $\nu$ instead of quadratic. 
More importantly, the $J$ damping decreases exponentially with filling $\bar n$, while
the $V$ attenuation effect {\em increases} with the filling.  
However, while $J/U$ will be necessarily 
small in the Mott phase (it is exponentially decreasing with increasing laser intensity), 
$V/U$ can be of order unity \cite{noteII}, so that the effect of $V$ can be very 
pronounced while that of $J$ will be subdominant, cf.\,Fig.\,\ref{finiteJ} below. 
For $U>V$, the period of the first full-amplitude coherence  
revival is $2\pi/V$ instead of $2\pi/U$. 
When $U/V$ is not a rational number, no full height revival occurs, and so-called fractional
revivals are obtained instead \cite{Robinett}. 

Due to the long-range nature and slow fall-off of the 
dipolar interaction with distance $\propto 1/r^3$, 
the next-nearest and next-to-next-nearest neighbors et cetera add other coupling terms of the form $\frac{V_{NN\mu}}2 \sum_{\{ij\}}^M \hat n_i \hat n_j $ to the Hamiltonian \eqref{H}, where $\{ij\}$ indicates a sum over next-nearest neighbors and $\lambda$ is their number,  $\sum_{\{ik\}} =\lambda$ at fixed $k$. Therefore, additional (slower) oscillation factors 
will be added in \eqref{UVrevival}. 
We take as a concrete example to illustrate this a 2D square lattice, with 
dipoles oriented perpendicular to the 2D plane.
This geometry implies that the closest next-nearest neighbors are a distance $\sqrt2 a$
from a given site, such that the associated coupling $V_{NN1}/V=1/\sqrt8$,
and the following next-to-next-nearest-neighbor sites at a distance $2a$, giving 
a coupling $V_{NN2}/V =1/8$.
Therefore, while the revival at integer multiples of $2\pi/V$ occurs with the full initial height when only nearest neighbors would be taken into account (i.e. for perfectly screened interactions at the NN distance), adding further next-nearest couplings, etc., makes the first full height revival peak occur much later.
The full height revivals related to truly long-range interactions thus occur at a very late and hence not experimentally observable stage after the sweep. The influence of the $V_{NN\mu}$ on the experimentally most relevant quantity, the 
attenuation of the first few revivals, is however small as long as $V/U\ll 1$. 
Our further discussion is based on the assumption that the Hamiltonian \eqref{H}
with only nearest-neighbor couplings describes the system, 
keeping in mind the above limitation.

\section{Modifying the Gutzwiller approach}  
Conventionally, the equations of motion resulting from \eqref{H} 
are solved (numerically)  with a Gutzwiller mean-field product state 
variational ansatz for the many-body wavefunction, cf., e.g. 
\cite{Kovrizhin,Natu}  
\bea
\ket{\rm GW} &=& \prod_{i=1}^M \sum_{n_i} f_i (n_i,t) \ket{n_i} \nn
&=& \sum_{n_1}\cdots \sum_{n_M} \left\{\prod_{i=1}^M  f_i (n_i,t)\right\} 
\ket{n_1,\ldots,n_M}
\label{GW}
\ea
with $f_i(n_i,t=0) = e^{-|\alpha_i^2|/2} {\alpha_i^{n_i}}/{\sqrt{n_i !}}$
for a coherent state \eqref{coh} 
and where $\ket{n_1,\ldots,n_M}= \ket{n_1} \otimes \cdots \ket{n_i}\cdots 
\otimes \ket{n_M}$ is the number basis state corresponding 
to noninteracting bosons occupying Wannier orbitals at each site.
The above variational wavefunction ansatz is a bosonic 
version of the fermionic product state wavefunction 
originally proposed by Gutzwiller \cite{Gutzwiller}. 

The mean-field theory implicit in the Gutzwiller ansatz for the wave function 
is expected to be exact for the standard onsite interaction Bose-Hubbard model when either $\bar n$ or $\nu$ \cite{Schuetzhold,SUF,Navez} are large; thus the
``mean-field parameter'' $\bar n\nu$, characterizing the validity of mean-field theory,  
should be sufficiently large. 
The Gutzwiller ansatz then becomes exact for contact interactions in infinite dimension \cite{Rokhsar}. Another extreme case of exact validity is a fully connected lattice \cite{Biroli}. 
The Gutzwiller wavefunction, in the ground state and for contact interactions, has  
been argued to be qualitatively correct for lower, and hence easily experimentally realized,
dimensions as well \cite{Krauth}. The Gutzwiller ansatz \eqref{GW} 
is however widely used for ground-state calculations also in the extended Bose-Hubbard model 
\cite{Trefzger,Goral,Scarola,Iskin}, where its rigorous validity for large mean-field
parameter $\bar n \nu$ is much less obvious than in the onsite interaction case. 
As we will now argue, the extended Bose-Hubbard Hamiltonian cannot be decoupled (in a nontrivial manner) into the sum of single-particle Hamiltonians, like the Hubbard model with contact interactions, where mean-field decoupling and the Gutzwiller ansatz have been shown to be equivalent \cite{Sheshadri}. 

\begin{center}
\begin{figure}[hbt]
\centering
\includegraphics[width=.44\textwidth]{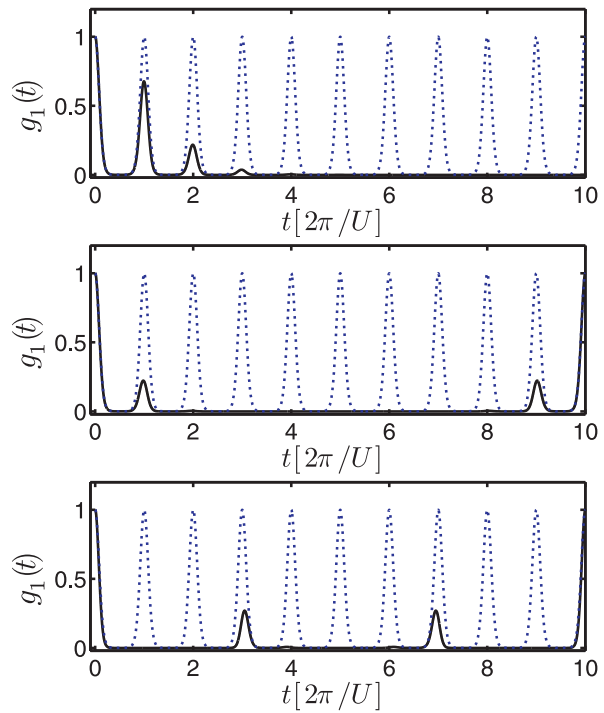}
\vspace*{0.5em}
\caption{\label{comparison} (Color online) Time evolution of $g_1$ with an initial coherent state 
for $V/U=0.05,0.1,0.3$, from top to bottom, with filling $\bar n =2$.
Solid line using the Gutzwiller variant \eqref{GWV}, dotted blue line
using the conventional Gutzwiller ansatz, Eq.\,\eqref{GW}.}
\end{figure}
\end{center} 
\vspace*{-2em}

The  mean-field decoupling for the tunneling term, in the homogeneous case, 
$\phi\equiv \langle \hat a_i\rangle$, leads from \eqref{H}, when $V=0$, to a sum of single-particle Hamiltonians, $\hat H_i = -J(\hat a^\dagger_i \phi + \phi^* \hat a_i -|\phi|^2)+\frac U2 \hat n_i( \hat n_i -1) $. On the other hand, the nonlocal interaction  term $\propto V$, after applying mean-field decoupling, just shifts the chemical potential, 
and thus remains dynamically ineffective. In particular, while the Gutzwiller
wavefunction \eqref{GW} reproduces the on-site-interaction modulations with period $2\pi/U$, 
no modulations of $g_1$ with period $2\pi/V$ are created after mean-field decoupling
has been performed.
Note that the (instantaneously) vanishing order parameter does not cause the failure of the conventional Gutzwiller wavefunction ansatz. For local interactions, collapse and
revival are captured accurately by the Gutzwiller wavefunction,  
even though the mean-field amplitude vanishes when first-order coherence collapses.
The inadequacy of the Gutzwiller ansatz for the 
time evolution of coherence in the extended Bose-Hubbard model thus is much stronger than that caused by the tunneling term $\propto J$ \cite{Fischer,Wolf}. The tunneling term 
can (at least in large dimensions) be accurately decoupled in mean-field, while this is not possible in a nontrivial way for the $V$ term in \eqref{H}, as argued above.  
We also observe in this regard that the expression \eqref{UVrevival} for the collapse and revival oscillations of $g_1$ is nonperturbative in either $1/\bar n$ or $1/(\bar n \nu)$, so that there is no obvious mean-field limit of the exact quantum evolution of first-order coherence in deep optical lattices. The quantum evolution of coherence can therefore serve as a sensitive probe of various time-dependent  ans\"atze
for the many-body wavefunction.

Based on 
the above reasoning, we devise a variant of the dynamical Gutzwiller wavefunction, which takes into account that the time evolution according 
to \eqref{H} significantly depends on the occupation of neighboring sites when $V$ is appreciably large. We assume still a product of variational amplitudes $f_i$ in the 
many-body wavefunction, but are now making the $f_i$ in addition 
dependent on the NN particle numbers with respect to the given site $i$, 
\bea
\ket{\rm GWV} 
= \sum_{n_1}\cdots \sum_{n_M} \left\{\prod_{i=1}^M  f_i (n_i,n_{[ji]},t)\right\} 
\ket{n_1,\ldots,n_M}\nn
\label{GWV}
\ea
where  $n_{[ji]}$ again indicates the set of all particle numbers on the NN sites $j$ to a given $i$, cf. Eq.\eqref{UVrevival}. 

\begin{center}
\begin{figure}[t]
\centering
\includegraphics[width=.45\textwidth]{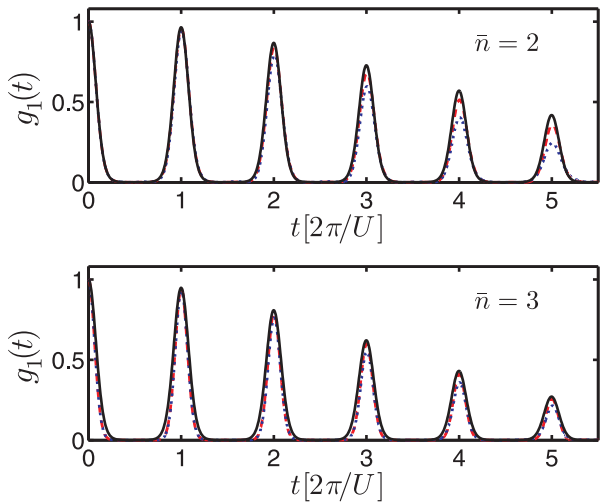}
\vspace*{0.5em}
\caption{\label{finiteJ} (Color online) Time evolution for fixed $V=0.015$ at two different
fillings $\bar n$ as indicated in the plots, 
and for varying initial tunneling coupling $J$, 
$J=0$ (solid line, analytical result), $J=0.01$ (red dashed)
and $J=0.02$ (blue dotted).
}
\end{figure}
\end{center} 
\vspace*{-2em}

The ansatz \eqref{GWV} is physically corresponding to the 
fact that the on-site distribution function $f_i$ should also depend on the particle numbers in neighboring sites for finite $V$. This can be directly seen from the equations of motion for the variational functions following from the extended Bose-Hubbard model \eqref{H} when $J=0$, see Eq.\,\eqref{GWV_eq} in the appendix
\begin{multline} 
i \frac{d f_{k}(n_{k}, n_{[jk]}, t)} {dt} \\
  = f_{k}(n_{k}, n_{[jk]}, t) n_{k}  
\left[ \frac{U}{2}(n_{k} -1) + \frac{V}{2} \sum_{[jk]}n_j
- \mu \right], \label{GWVEOM}
\end{multline}
where $\mu$ is the chemical potential. 
Solving this equation generates the dynamical phase factors in \eqref{UVrevival}. 
We have thus verified 
that as a benchmark for the validity of \eqref{GWV} the analytical result \eqref{UVrevival} is reproduced and hence the exact quantum evolution in the limit of zero tunneling.
We illustrate this by comparing \eqref{UVrevival} with the evolution using either
\eqref{GW} or \eqref{GWV}, 
see Fig.\ref{comparison}. 
The evolution according to \eqref{GWV} is identical with \eqref{UVrevival}, 
while propagating the initial coherent state with the conventional Gutzwiller ansatz \eqref{GW} shows no modulation of the revival peaks with period $2\pi/V$ at all. 

We anticipate that the Gutzwiller variant (GWV) 
ansatz \eqref{GWV} will also be a good approximation of the dynamical many-body state for a given $U,V$ when the tunneling rate $J$ is sufficiently small.
We illustrate the effect of finite $J$ by numerically computing in a 1D lattice 
the time evolution of an initially coherent state, using the GWV tunneling energy functional \eqref{EiJ} derived in the appendix. We show the results in Fig.\,\ref{finiteJ}.
There is the expected (real) damping occurring for finite $J$ in addition to the 
attenuation of the first few coherence revivals obtained by finite $V$.  
When $\bar n V\gtrsim J$ (assuming $U\ge V$), which is the regime of validity for the ansatz \eqref{GWV}, the $V$ attenuation will dominate the $J$ damping;
cf.\,Fig.\,\ref{finiteJ}. 
Conversely, for $J\gtrsim \bar nV$, the conventional Gutzwiller ansatz \eqref{GW} will again be applicable, i.e., yield in this regime a better approximation than the GWV ansatz \eqref{GWV}. Finally, coherent state amplitudes \eqref{coh} apply in \eqref{GW} when one is entering deep into the superfluid phase, which has $J\gg \bar nU,\bar n V$.

\section{Number squeezed initial state}

To employ our variant of the Gutzwiller ansatz for a less simple initial state than the 
coherent state \eqref{coh}, we now investigate the 
NN coupling effect on collapse and revival when 
finite number squeezing is present in the initial state. 
We take as the initial state the numerically determined 
ground state of the Hamiltonian \eqref{H} in one dimension 
for finite $J/U$ and small system sizes ($M=7$), by exact diagonalization (ED), 
so that the initial state is determined as the ground state
of \eqref{H} for a number-squeezed state. 

We evolve this number-squeezed initial state within the  
GWV wavefunction ansatz \eqref{GWV} in a 1D lattice.
To implement GWV propagation, 
we proceed by the following prescription. We convert the 
number basis amplitudes in the general many-body wavefunction obtained from ED,  
$|\Psi\rangle = \sum_{n_{1}, \ldots, n_{M}} f(n_{1}, \ldots, n_{M}) |n_{1}, \ldots, n_{M} \rangle$
by the prescription 
$f_{i}(n_{i}) = \sqrt{\sum_{n_{1},\ldots, n_{i-1}, n_{i+1}, \ldots, n_{M}} |f (n_{1}, \ldots, n_{i}, \ldots, n_{M})|^{2}}$
into real amplitudes of an {\em initial} product state of the GW type, 
and then propagate this initial state by the equations of motion of the GWV wavefunction, 
Eq.\,\eqref{GWVEOM}. A comparison we made with the ED propagation of the full initial data 
shows that 
similar results are obtained 
as long as $\eta= U/({\nu J})$ does not become too large [i.e., larger than $\ord(\bar n)$], in which case both initial coherences $g_1(0)$ are sufficiently 
close to the coherent state value \cite{comparison}. 
%

\begin{center}
\begin{figure}[t]
\centering
\includegraphics[width=.35\textwidth]{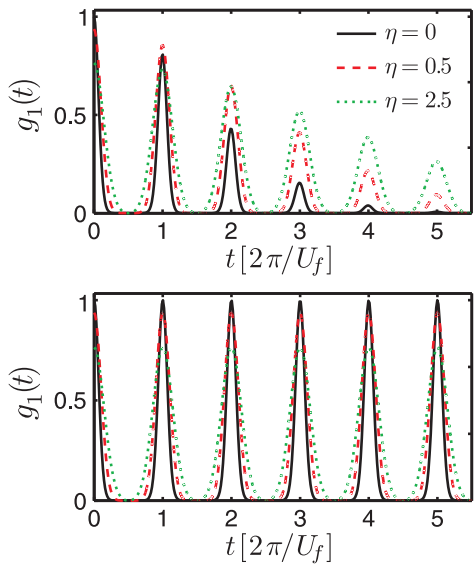}
\vspace*{0.5em}
\caption{\label{finite_eta}  
(Color online) Top: Time evolution of $g_1$ according to the GWV wavefunction \eqref{GWV} 
for number-squeezed initial states with squeezing parameter $\eta=U_i/(\nu J)$
as indicated in the plot ($\nu=2,\bar n =3, |i-j|=3$). 
The initial and final $V=0.03$, final $U=U_f=1$, as well as initial $J=1$ and final $J=0$ are kept fixed, while $\eta$ is varied by changing the initial coupling $U=U_i$. 
Bottom: Time evolution for the 
GW ansatz \eqref{GW}, with identical parameter set
and the same initial data.}
\end{figure}
\end{center} 
\vspace*{-2em}

The basic result we obtain, upon increasing the {\em squeezing parameter} 
$\eta $
from zero (corresponding to a coherent state), 
is that revival peaks broaden but also that at the same time the 
attenuation of the first few $U$-revivals is diminished, see Fig.\,\ref{finite_eta} \cite{noteIII}. 
This is related to the fact that the narrower the (initial) distribution, 
the less phase spread in the dynamical phase factor 
is generated  by the many-site coupling due to the 
$V$ term in \eqref{H}. In fact, the same broadening is obtained with the conventional 
Gutzwiller ansatz \eqref{GW}, but like for the coherent initial state ($\eta =0$), 
no attenuation of the revival peak height is observed in distinction to the GWV evolution \eqref{GWV}, cf.\,Fig.\,\,\ref{finite_eta}. 
We note, in particular, that the Gutzwiller approach also fails to reproduce that, for larger $\eta$, the attenuation of the first revival peak heights relative to the initial coherence signal is reduced. 


Collapse and revival with reduced initial amplitude but less attenuation thus 
persists for not too large squeezing parameter $\eta$. 
This opens up the possibility of probing the existence and time development of 
coherence revivals, and thus the interplay of quantum fluctuations and order parameter value, 
for various initial data sampled from around the transition point
between superfluid and localized phases.
The collapse and revival phenomenon ultimately 
owes its existence to the granularity of matter, i.e. the fact that particle number is 
quantized in integer multiples. Measuring the collapse and revival oscillations around the 
transition point with single-atom resolution \cite{Bakr}  
should therefore afford interesting insights
into the microscopic physics, that is the many-body
wavefunction describing the transition.


\section{conclusion}
The collapse and revival oscillations of first-order coherence were shown to be 
a sensitive measure of  the NN couplings in the extended Bose-Hubbard model.  
After a rapid quench from the superfluid side, the first few revival peaks are significantly
attenuated for moderate NN couplings. 
Even relatively small dipole moments of molecules stored in optical lattices 
can thus be detected by the attenuation of the height of the first few 
revival peaks. The attenuation effect is strong enough  
to overcome masking inhomogenity effects like those caused by the
external harmonic trapping, in distinction to the weak real damping 
caused by residual tunneling between deep wells in the localized phases.  
The Gutzwiller variant ansatz we determined from 
the exact collapse and revival quantum evolution in the 
limit of zero tunneling is also potentially relevant for exploring the 
phase boundaries in the ground state of the extended Bose-Hubbard model. A particular example are supersolid phases in the limit of large NN coupling $V$, where a comparison 
should be instructive to those obtained by the standard Gutzwiller approach \cite{Iskin}.

We expect the morphology of quantum phase revivals to 
be a valuable future tool for both the quantitative
analysis of interaction couplings of Hamiltonians of atoms/molecules 
with long-range interactions stored in deep optical lattices, as well 
as a sensitive test-bed for the validity of various approximation schemes 
for the many-body wavefunction.

Finally, a possible extension of the present work is to include pair-exchange terms 
between lattice sites, i.e., by adding a contribution $\hat H_{\rm pair} \propto 
\sum_{<ij>}^M (\hat a_i^\dagger)^2 \hat a_j^2$ to the Hubbard Hamiltonian \eqref{H}, see e.g.\,\cite{Illuminati}, from which novel interaction-induced effects on the dynamical evolution of first- and second-order coherence can be expected \cite{Xiong}.

\acknowledgments
This research was supported by the 
Brain Korea BK21 program and the 
NRF of Korea, grant No. 2010-0013103. 

\appendix*
\section{Equations of motion from the GWV ansatz}
We delineate here the derivation of the equations of motion for the 
GWV wavefunction \eqref{GWV}. For clarity and simplicity of notation,
we restrict ourselves to a one-dimensional lattice; generalizations to 
higher lattice dimensions 
are straightforward. 
 
We deduce the equations of motion variationally, starting from the functional 
\begin{equation} 
\mathcal{E} = \bra{\rm GWV}  i \partial_{t} - \sum_{i} \left\{\hat{H}_i
- \mu \hat{n}_{i}\right\} \ket {\rm GWV}, \label{E}
\end{equation}
where $\hat{H}_i = -J \sum_{j=i\pm 1}  
\hat a^\dagger_i \hat a_j +
\frac{U}{2}\hat{n}_{i} (\hat{n}_{i} - 1) 
+ \frac{V}{2} \hat{n}_{i} \hat{n}_{i+1}
+\frac{V}{2} \hat{n}_{i} \hat{n}_{i-1}$. 

We first consider the case with no tunneling, $J=0$, for which \eqref{GWV} 
reproduces the exact quantum evolution. 
Due to the fact that the GWV wavefunction \eqref{GWV} is still 
a product state ansatz like the original Gutzwiller wavefunction, the energy 
functional \eqref{E} decomposes into contributions of a given site: 
\begin{widetext}
\begin{equation} \label{Ei}
         \begin{split}
            \mathcal{E}_{i} & = \sum_{n_{i}, n_{i\pm 1}, \cdots} f_{i}^{*}(n_{i}, n_{i+1}, n_{i-1}, t) f_{i-1}^{*}(n_{i-1}, n_{i}, n_{i-2}, t) f_{i+1}^{*}(n_{i+1}, n_{i+2}, n_{i}, t)\cdots \\ 
                            & \times\left[ i\partial_{t}^{(i)} - H_{i}(n_{i}, n_{i+1}, n_{i-1}) + \mu n_{i} \right] f_{i}(n_{i}, n_{i+1}, n_{i-1}, t) f_{i-1}(n_{i-1}, n_{i}, n_{i-2}, t) 
                            f_{i+1}(n_{i+1}, n_{i+2}, n_{i}, t)\cdots 
         \end{split}
      \end{equation}
                \end{widetext}     
where $i\partial_{t}^{(i)}$ now acts on $f_{i}(n_{i}, n_{i+1}, n_{i-1}, t)$ only. 
Furthermore, when $J=0$, the phase factors of e.g. the functions
$f_{i+1}(n_{i+1}, n_{i+2}, n_{i}, t)$ 
and $f_{i-1}(n_{i-1}, n_{i}, n_{i-2}, t)$, 
containing all their time and NN particle number dependence, cancel with those of their
conjugates in ${\cal E}_i$, so that they can be replaced by the onsite distributions 
of the initial state, $ f_{i\pm 1}(n_{i\pm 1}, n_{i}, n_{i\pm 2},t) \rightarrow f^0_{i\pm 1} (n_{i\pm 1})$. 
The equations of motion then are
\begin{equation} \label{GWV_eq}
         [i\partial_{t}^{(i)} - H_{i}(n_{i}, n_{i+1}, n_{i-1}, t) + \mu n_{i}] f_{i}(n_{i}, n_{i+1}, n_{i-1}, t) = 0, 
      \end{equation} 
      with the normalization condition $\sum_{n_{i}} |f_{i} (n_{i}, n_{i+1}, n_{i-1}, t)|^{2} = 1$, leading to number conservation in the form 
     $\sum_{n_{i}} n_i |f_{i} (n_{i}, n_{i+1}, n_{i-1}, t)|^{2} = \bar n$.
The above equations of motion give the exact time evolution when $J=0$.  

Due to $[ \hat{H} (\hat{n}_{i}, \hat{n}_{i\pm 1}),\hat g (\hat n_i, \hat n_{i\pm 1}) ]=0$,  where $\hat g(\hat n_i, \hat n_{i\pm 1})$ is an arbitrary operator function, 
a number of additional conserved quantities can be constructed.
One such quantity is, choosing $ \hat g(\hat n_i, \hat n_{i\pm 1}) 
= \hat n_{i} + \hat n_{i+1} + \hat n_{i-1}$, 
\begin{equation} \label{conserved_C} 
         \begin{split} 
            & \bra{\rm GWV} \hat n_{i} + \hat n_{i+1} + \hat n_{i-1} \ket{\rm GWV}
         \\ & = \sum_{n_{i}, n_{i\pm 1}} (n_{i} + n_{i+1} + n_{i-1}) |{F}_{i} 
            |^{2} 
= \mathcal{C}, 
         \end{split}   
      \end{equation} 
where the threefold product amplitude $F_i$ is defined by 
\begin{equation}
\begin{split}
& F_i (n_i,n_{i+1},n_{i-1},t)) \\
& \equiv 
 f_i(n_i,n_{i+1},n_{i-1},t)\tilde f_{i+1}(n_{i+1},n_i,t) \tilde f_{i-1}(n_{i-1},n_i,t)
\label{defFi}
 \end{split} 
\end{equation}
$[= f_i(n_i,n_{i+1},n_{i-1},t)f^0_{i-1}(n_{i-1})f^0_{i+1}(n_{i+1})$ for $J=0$]. 
Here, the $\tilde f_{i\pm 1}(n_{i\pm 1},n_i,t)$ are defined by stripping off the 
$n_{i\pm 2}$ dependence from the full $f_{i\pm 1}$ amplitudes, e.g.
$f_{i-1}(n_{i-1},n_i,n_{i-2},t)=\tilde f_{i-1}(n_{i-1},n_i,t)\exp[-i\frac V2 n_{i-1}n_{i-2} t]$.
This particular choice of $\hat g$ and the definition of $F_i$ in the form above
is motivated by $\cal C$ being conserved also in the $J\neq 0$ case (see below).  
In a lattice of any dimension and geometry, $\mathcal{C} = (\nu+1 )\bar n$; 
for the 1D lattice $\mathcal{C} = 3\bar n$. 
    
When $J$ does not vanish and is small, the ansatz \eqref{GWV} can be applied to describe temporal evolution of the many-body state under certain conditions. Primarily, the latter consist in the requirement that the {modulus} of the variational amplitudes is not appreciably changed by finite and small tunneling and still dominated by the initial distribution, $f_i^0(n_i)$, as well as that phase factors are to lowest order still given by the $J=0$ expression, 
$\exp\left[-i \frac V2(n_i n_{i+1} +n_i n_{i-1})t - i \frac U2 n_i(n_i-1)t \right]$. 

The onsite contribution of the tunneling term ${\mathcal E}_{i,J}$  to the functional 
\eqref{E} consists of four terms assigned with $\hat a^\dagger_i \hat a_{i\pm 1}$
and $\hat a^\dagger_{i\pm 1} \hat a_i$. 
Consider one such term, $-J\bra{\rm GWV} \hat{a}_{i+1}^{\dag} \hat{a}_{i} \ket{\rm GWV}$.
Under the above assumption, to lowest order in $J/V$ 
(assuming $U\ge V$), 
it can be shown to take the approximate form 
\begin{widetext}
\begin{equation}
-J  \langle \hat{a}_{i+1}^{\dag} \hat{a}_{i} \rangle 
\simeq 
-J \sum_{n_{i}, n_{i\pm1}} \sqrt{(n_{i}+1)n_{i+1}} F^*_i (n_{i}, n_{i+1}, n_{i-1},t) 
F_i(n_i+1,n_{i+1}-1,n_{i-1},t)
\sum_{n_{i+2}} |f_{i+2}(n_{i+2})|^{2} \mathrm{exp}\left(iV n_{i+2} t\right),
\end{equation}
\end{widetext} 
with similar expressions for the other three terms. 
In the short time domain $t\ll 2\pi/V$ we are interested in [cf. the final paragraph
of section II], the oscillating factor at the end of the right-hand side can be omitted, so that collecting
all four terms, the tunneling contribution to the functionals 
\eqref{E} respectively \eqref{Ei} becomes 
\begin{widetext}
\begin{equation} 
       \begin{split}
            \mathcal{E}_{i,J} & \simeq J \sum_{n_{i}, n_{i\pm1}} {F}_{i}^{*}(n_{i}, n_{i+1}, n_{i-1}, t)\left[ \sqrt{n_{i} (n_{i+1} + 1)} {F}_{i}(n_{i} - 1, n_{i+1} + 1, n_{i-1}, t) \right.\\
            & \left. + \sqrt{n_{i} (n_{i-1} + 1)} {F}_{i}(n_{i} - 1, n_{i+1}, n_{i-1}+1, t) + \sqrt{(n_{i}+1) n_{i+1}} {F}_{i}(n_{i} + 1, n_{i+1}-1, n_{i-1}, t)  \right.
            \\
                              &\left.
            + \sqrt{(n_{i}+1) n_{i-1}} {F}_{i}(n_{i} + 1, n_{i+1}, n_{i-1}-1, t)  \right] , 
                   \end{split}
                   \label{EiJ}
    \end{equation}      
    \end{widetext} 
where the normalization condition now reads 
$\sum_{n_{i},n_{i\pm 1}} |F_{i} (n_{i}, n_{i\pm 1}, t)|^{2} = 1$. Writing 
the interaction part as well in the product amplitudes \eqref{defFi}, the equations
of motion can then be derived from the functional derivative $\delta \mathcal{E}_{i}/\delta {F}_{i}^{*} (n_{i}, n_{i+1}, n_{i-1}, t)=0$. 
    
Finally, one can easily verify that the tunneling term preserves the conservation of the sum of onsite and NN numbers,   $[- J (\hat{a}^\dagger_{i}\hat{a}_{i+1} + \hat{a}_{i}^{\dag}\hat{a}_{i- 1}+\hat a^\dagger_{i+1} \hat a_i + \hat a^\dagger_{i-1} \hat a_i) , \hat{n}_{i} +  \hat{n}_{i+ 1} +\hat{n}_{i-1}] = 0,$ so that the conservation law \eqref{conserved_C} still holds, providing a benchmark for the accuracy
of numerical calculations.


\begin{thebibliography}{9999}

\newfont{\cyr}{wncyr10}

\bibitem{Gutzwiller} M.\,C. Gutzwiller, 
Phys. Rev. Lett. {\bf 10}, 159 (1963). 

\bibitem{Hubbard} J. Hubbard,
Proc. Roy. Soc. (London)  A {\bf 276}, 238 (1963).  


\bibitem{Kanamori} J. Kanamori, 
Prog. Theor. Phys. {\bf 30}, 275 (1963). 

\bibitem{Jaksch} D.~Jaksch, C.~Bruder, J.\,I.~Cirac, C.\,W.~Gardiner, and P.~Zoller,
Phys.~Rev.~Lett.~{\bf 81}, 3108 (1998). 

\bibitem{Trefzger} A recent  overview 
is contained in 
C. Trefzger, C. Menotti, B. Capogrosso-Sansone, and
M. Lewenstein, 
J. Phys. B: At. Mol. Opt. Phys. {\bf 44}, 193001 (2011).


\bibitem{Baranov} M. Baranov, 
Phys. Rep. {\bf 464}, 71 (2008). 

\bibitem{history} We note here that the ``extended" conventionally attached to the 
beyond on-site-interaction Hubbard model is to some extent 
a historical misnomer: Already in \cite{Hubbard} it is pointed out 
that the $V$ term in \eqref{H} can be appreciable even with 
screened Coulomb interactions. 

\bibitem{Robinett} R.\,W. Robinett, 
Phys. Rep. {\bf 392}, 1 (2004). 

\bibitem{Greiner} M. Greiner, O. Mandel, T.\,W. H\"ansch,
and I. Bloch, 
Nature {\bf 419}, 51 (2002).

\bibitem{Will} S. Will, 
T. Best, U. Schneider, L. Hackerm\"uller, D.-S. L\"uhmann, and I. Bloch, 
Nature {\bf 465}, 197 (2010).

\bibitem{Dziarmaga}  J. Dziarmaga, Adv. Phys. {\bf 59}, 1063 (2010).    

\bibitem{Bakr} W.\,S. Bakr, 
A. Peng, M. E. Tai, R. Ma, J. Simon, J.\,I. Gillen, S. F\"olling, L. Pollet, and M. Greiner, 
Science {\bf 329}, 547 (2010).

\bibitem{Weitenberg} 
C. Weitenberg, 
M. Endres, J.\,F. Sherson, M. Cheneau, P. Schau\ss, T. Fukuhara, I. Bloch, and S. Kuhr,
Nature {\bf 471}, 319 (2011). 

\bibitem{Wright} E.\,M. Wright, D.\,F. Walls, and J.\,C. Garrison,
Phys. Rev. Lett. {\bf 77}, 2158 (1996). 

\bibitem{You} A. Imamo\=glu, M. Lewenstein, and L. You,
Phys. Rev. Lett. {\bf 78}, 2511 (1997).  

\bibitem{Fischer} U.\,R. Fischer and R. Sch\"utzhold, Phys. Rev. A {\bf 78},
061603(R) (2008).

\bibitem{Wolf}  F.\,A. Wolf, I. Hen, and M. Rigol, Phys. Rev. A {\bf 82},
043601 (2010).

\bibitem{Schachenmayer} J. Schachenmayer, A.\,J. Daley, and P. Zoller, 	
Phys. Rev. A {\bf 83}, 043614 (2011).

\bibitem{Johnson} 	P.\,R. Johnson, E. Tiesinga, J.\,V. Porto, and C.\,J. Williams, 
New J. Phys. {\bf 11}, 093022 (2009).

\bibitem{Tiesinga}  E. Tiesinga and P. R. Johnson, 	
Phys. Rev. A {\bf 83}, 063609 (2011).  

\bibitem{Schuetzhold} U.\,R. Fischer, R. Sch\"utzhold, and M. Uhlmann, Phys. Rev. A {\bf 77}, 043615 (2008); R. Sch\"utzhold, 
M. Uhlmann, Y. Xu, and U.\,R. Fischer,
Phys. Rev. Lett. {\bf 97}, 200601  (2006). 

\bibitem{Buchhold} M. Buchhold, U. Bissbort, S. Will, and W. Hofstetter,
Phys. Rev. A {\bf 84}, 023631 (2011). 

\bibitem{noteI} We use the term ``attenuation'' to distinguish the reduction in amplitude
of the first few coherence revivals by NN coupling from the real damping caused, e.g., by
the tunneling coupling between sites \cite{Fischer}. 

\bibitem{Goral} K. G\'oral, L. Santos, and M. Lewenstein,
Phys. Rev. Lett. {\bf 88}, 170406 (2002). 

\bibitem{Scarola} V.\,W. Scarola and S. Das Sarma, 
Phys. Rev. Lett. {\bf 95}, 033003 (2005). 

\bibitem{Iskin} M. Iskin,   
Phys. Rev. A {\bf 83}, 051606(R) (2011).

\bibitem{noteII} For large dipolar interactions and spherical onsite trapping potential (for which the onsite contribution of dipolar interaction averages to zero) $V/U = \ord [g_d\sigma^3/(g a^3)]$, approximating the Wannier states by Gaussians of width $\sigma$; $a$ is a 1D lattice constant and $g,g_d$ are contact and dipole-dipole coupling constants, respectively. 

\bibitem{Kovrizhin} D.\,L. Kovrizhin, G. Venketeswara Pai, and S. Sinha,
Europhys. Lett. {\bf 72}, 162 (2005). 

\bibitem{Natu} S.\,S. Natu, K.\,R.\,A. Hazzard, and E.\,J. Mueller,
Phys. Rev. Lett. {\bf 106}, 125301 (2011). 

\bibitem{SUF} R. Sch\"utzhold, M. Uhlmann, and U.\,R. Fischer,
Phys. Rev. A {\bf 78}, 033604 (2008). 

\bibitem{Navez} P. Navez and R. Sch\"utzhold, Phys. Rev. A {\bf 82}, 063603 (2010). 

\bibitem{Rokhsar}  D.\,S. Rokhsar and B.\,G. Kotliar, 
Phys. Rev. B {\bf 44}, 10328 (1991). 

\bibitem{Biroli} B. Sciolla and G. Biroli, 
Phys. Rev. Lett. {\bf 105}, 220401 (2010).  


\bibitem{Krauth}  W. Krauth, M. Caffarel, and J.-P. Bouchaud,
Phys. Rev. B {\bf 45}, 3137 (1992). 


\bibitem{Sheshadri} K. Sheshadri, H.\,R. Krishnamurthy, R. Pandit,  
and T.\,V. Ramakrishnan, 
Europhys. Lett. {\bf 22}, 257 (1993). 

\bibitem{comparison} For the $\eta =2.5$ data shown (Fig.\,\ref{finite_eta} top), the difference in the peak height of the fifth revival amounts to 11 \%.

\bibitem{noteIII} 
We choose to vary the initial $U$ 
to increase $\eta$, because decreasing $J$ entails changing another squeezing parameter related to off-site coupling,  $V/J$, as well. We have 
verified that the (more easily experimentally accessible) latter option of changing $J$ 
yields a negligible difference in the correlation signal at the same $\eta$ as long as $V\ll U$.



\bibitem{Illuminati}  G. Mazzarella, S.\,M. Giampaolo, and F. Illuminati, 
Phys. Rev. A {\bf 73}, 013625 (2006).  

\bibitem{Xiong} U.\,R. Fischer, K.-S. Lee, and B. Xiong, 
Phys. Rev. A {\bf 84}, 011604(R) (2011). 

\end{thebibliography}
\end{document}